\documentclass[aps,preprint,nofootinbib]{revtex4}
\usepackage{amsfonts}
\usepackage{amsmath}
\usepackage{amssymb}
\usepackage{graphicx}%
\providecommand{\U}[1]{\protect\rule{.1in}{.1in}}

\begin{document}
\preprint{ }
\title[ ]{Annihilation of the electron-positron pairs in the positronium ion
Ps$^-$ and bi-positronium Ps$_2$.}
\author{Alexei M. Frolov}
\email{afrolov@uwo.ca}
\affiliation{Department of Chemistry, University of Western Ontario, London,
Canada}
\vspace{0.5cm}

\begin{abstract}
Rates of the two-, three-, four- and five-photon annihilations of the
electron-positron pairs are determined numerically for the three-body
positronium ion Ps$^-$ ($e^- e^+ e^-$) and four-body bi-positronium
`molecule' Ps$_2$ ($e^- e^+ e^- e^+$). The values obtained in our
computations are $\Gamma_{2 \gamma}($Ps$^-) \approx$ 2.08048530525$\cdot
10^{9}$ $sec^{-1}$, $\Gamma_{3 \gamma}($Ps$^-) \approx$ 5.6364151550$\cdot
10^{6}$ $sec^{-1}$, $\Gamma_{4 \gamma}($Ps$^-) \approx$ 3.075$\cdot 10^{3}$
$sec^{-1}$, $\Gamma_{5 \gamma}($Ps$^-) \approx 5.383$ $sec^{-1}$ and
$\Gamma_{2 \gamma}($Ps$_2) \approx$ 4.4385952$\cdot 10^{9}$ $sec^{-1}$,
$\Gamma_{3 \gamma}($Ps$_2) \approx$ 1.202497$ \cdot 10^7$ $sec^{-1}$,
$\Gamma_{4 \gamma}($Ps$_2) \approx$ 6.562$\cdot 10^3$ $sec^{-1}$, $\Gamma_{5
\gamma}($Ps$_2) \approx$ 11.484 $sec^{-1}$. The four- and five-photon
annihilation rates are significantly smaller than the corresponding two- and
three-photon annihilation rates known for these systems. We also determine
the rates of one- and zero-photon annihilation for the Ps$^{-}$ ion and
Ps$_2$ system. The corresponding numerical values are $\Gamma_{1
\gamma}($Ps$^-) \approx$ 3.82491$\cdot 10^{-2}$ $sec^{-1}$,
$\Gamma_{1 \gamma}($Ps$_2) \approx$ 1.94188$\cdot 10^{-1}$ $sec^{-1}$ and
$\Gamma_{0 \gamma}($Ps$_2) \approx$ 2.32197$\cdot 10^{-9}$ $sec^{-1}$.\\
Pusblished almost `as is' in Phys. Rev. A {\bf 80}, 014502 (2009).
\end{abstract}
\volumeyear{ }
\volumenumber{ }
\issuenumber{ }
\eid{ }
\maketitle

\newpage

\section{Introduction}

In this work we consider the annihilation of electron-positron pairs (or
$(e^-, e^+)-$pairs, for short) in the three-body positronium ion Ps$^-$
and four-body bi-positronium system (or molecule) Ps$_2$. Our main goal is
to evaluate the four- and five-photon annihilation rates \cite{PRA83} in
these systems. Another goal is to re-evaluate the `traditional' two- and
three-photon annihilation rates known for these two systems. In addition to
these values we determine the one- and zero-photon annihilation rates which
are also of interest in some applications. In general, the $n-$photon
annihilation of the electron-positron pair is written in the form
\begin{equation}
 e^{-} + e^{+} = \hbar \omega_1 + \hbar \omega_2 + \hbar \omega_3 + \ldots
 + \hbar \omega_n \label{eq1}
\end{equation}
where $n$ is the total number of the emitted photons, $\omega_i$ ($i = 1,
\ldots, n$) are the corresponding photon frequencies and $\hbar \approx
1.054571628 \cdot 10^{-34}$ $J \cdot sec$ is the Planck constant (also
called the reduced Planck constant or Dirac constant). Annihilation of
electron-positron pairs from bound states of various polyelectrons is
determined by the total annihilation rate $\Gamma$ which is the sum of
partial annihilation rates $\Gamma_{n \gamma}$, i.e. $\Gamma = \Gamma_{2
\gamma} + \Gamma_{3 \gamma} + \Gamma_{4 \gamma} + \Gamma_{5 \gamma} +
\ldots$, where $\Gamma_{n \gamma}$ is the partial $n-$photon annihilation
rate. For some additional conditions (see below), the one-photon
annihilation and zero-photon annihilation of the electron-positron pair are
also possible. For different polyelectrons, different atoms and molecules
which contain positrons the numerical values of $\Gamma, \Gamma_{2 \gamma},
\Gamma_{3 \gamma}, \Gamma_{4 \gamma}$, etc can be substantially different.
Below, we consider the annihilation of electron-positron pairs in the
three-body Ps$^-$ ion and four-body Ps$_2$ system (bi-positronium).

\section{Annihilation of the positronium ion}

Positron annihilation in the three-body Ps$^-$ ion was considered in a
number of studies (see, e.g., \cite{BD1}, \cite{Ho1}, \cite{Fro05} and
references therein). Note that the Ps$^-$ ion (or $e^{-} e^{+} e^{-}$ ion)
has only one bound state which is the singlet (ground) $1^1S(L = 0)-$state.
It follows from here that in the Ps$^-$ ion one electron-positron pair is
always in its singlet ${}^1S$-state, while second such a pair is in its
triplet ${}^3S$-state. Therefore, annihilation of the ($e^{-},e^{+}$)-pair
in the Ps$^-$ ion may proceed with the emission of an arbitrary (even or
odd) number of photons. In reality, such an annihilation proceeds with the
emission of the one-, two-, three- or more photons. The one-photon
annihilation rate (or width) $\Gamma_{1 \gamma}$ is written in the following
form \cite{Kru} (see also \cite{Fro05})
\begin{eqnarray}
 \Gamma_{1 \gamma} = \frac{64 \pi^2}{27} \cdot \alpha^8 \cdot c \cdot
 a_0^{-1} \cdot \langle \delta_{321} \rangle = 1065.7569198 \cdot \langle
 \delta_{321} \rangle  \; \; \; sec^{-1} \; \; \; , \nonumber
\end{eqnarray}
where $\alpha = 0.7297352568 \cdot 10^{-2}$ is the fine structure constant,
$c = 0.299792458 \cdot 10^{9}$ $m \cdot sec^{-1}$ is the velocity of
light, and the Bohr radius $a_0$ equals $0.5291772108 \cdot 10^{-10}$ $m$
\cite{COD1}. The value $\langle \delta_{321} \rangle$ used in this formula
is the expectation value of the triple delta-function computed for the
ground state of the Ps$^-$ ion. This value is the probability of finding
all three-particles at one spatial point. In actual computations the triple
delta-function $\delta_{321}$ is computed with the use of the following
formulas $\delta_{321} = \delta({\bf r}_{32}) \delta({\bf r}_{21}) =
\delta({\bf r}_{31}) \delta({\bf r}_{21}) = \delta({\bf r}_{31}) \delta({\bf
r}_{32})$. In some works, the triple delta-function for the Ps$^-$ ion is
designated as $\delta_{+--}$. By using the expectation value of the triple
delta-function $\langle \delta_{321} \rangle$ for the Ps$^-$ ion from
\cite{Fro07a} one finds from the formula given above that $\Gamma_{1
\gamma} \approx 3.82340 \cdot 10^{-2}$ $sec^{-1}$. Note that the total
non-relativistic energy of the ground $1^1S-$state of the Ps$^{-}$ ion
obtained with the same wave function is -0.26200 50702 32980 10777 03745
$a.u.$ It is lowest variational energy to-date.

Consider now the two- and three-photon annihilation rates in the Ps$^-$ ion.
For an arbitrary atom/molecule which contains electrons and positrons the
rates of two- and three-photon annihilation of the $(e^-,e^+)-$pair are
\cite{Fro07a}
\begin{eqnarray}
 \Gamma_{2 \gamma} = 4 \pi \alpha^4 c a^{-1}_0 \Bigl[ 1 - \frac{\alpha}{\pi}
 \Bigl( 5 - \frac{\pi^2}{4} \Bigr)\Bigr] \langle \delta({\bf r}_{+-})
 \rangle \approx 4 \cdot 50.17280269804 \cdot 10^{9} \cdot \langle
 \delta_{+-} \rangle \; sec^{-1} \label{ep2}
\end{eqnarray}
and
\begin{eqnarray}
 \Gamma_{3 \gamma} = \frac{16 (\pi^2 - 9)}{9} \alpha^5 c a^{-1}_0 \langle
 \delta({\bf r}_{+-}) \rangle \approx \frac43 \cdot 1.35927229774 \cdot
 10^8 \langle \delta_{+-} \rangle \; sec^{-1} \; \; \; , \label{ep3}
\end{eqnarray}
respectively. Both these formulae explicitly contain the expectation value
of the two-body electron-positron delta-function $\delta({\bf r}_{+-}) =
\delta_{+-}$. The expression for $\Gamma_{2 \gamma}$, Eq.(\ref{ep2}), also
includes the lowest order radiative correction \cite{Harr}. Note again that
the two-photon annihilation of the $(e^-, e^+)-$pair proceeds only from its
singlet states, while analogous three-photon annihilation is possible only
from the triplet state of the electron-positron pair.

In actual applications to polyelectrons the expressions, Eqs.(\ref{ep2}) -
(\ref{ep3}), must be multiplied by the total number of the singlet/triplet
electron-positron pairs ($n$) and corresponding statistical weights of the
considered singlet/triplet spin states. For the considered Ps$^-$ ion we
have $n = 2$, while statistical weights of the singlet and triplet states
equal $\frac14$ and $\frac34$, respectively. Therefore, from the formulae
presented above one finds
\begin{eqnarray}
 \Gamma_{2 \gamma}({\rm Ps}) = n \pi \alpha^4 c a^{-1}_0 \Bigl[ 1 -
 \frac{\alpha}{\pi} \Bigl( 5 - \frac{\pi^2}{4} \Bigr)\Bigr] \langle
 \delta({\bf r}_{+-}) \rangle \approx 100.3456053781 \cdot 10^{9} \langle
 \delta_{+-} \rangle \; sec^{-1} \label{An2g}
\end{eqnarray}
and
\begin{eqnarray}
 \Gamma_{3 \gamma}({\rm Ps}^-) = n \frac{4 (\pi^2 - 9)}{3} \alpha^5 c
 a^{-1}_0 \langle \delta({\bf r}_{+-}) \rangle \approx 2.718545954 \cdot
 10^8 \langle \delta_{+-} \rangle \; sec^{-1} \; \; \; ,
\end{eqnarray}
where $\langle \delta_{+-} \rangle$ is the expectation value of the
electron-positron delta-function determined for the $1^1S$-state in the
Ps$^-$ ion.

Now, let us discuss the multiphoton annihilation of the electron-positron
pairs in the Ps$^-$ ion. The multiphoton annihilation includes, in
particular, the cases of four- and five-photon annihilation. It was shown in
\cite{PRA83} that the rates of the four- and two-photon annihilation in the
para-positronium (i.e. in the $(e^{-}, e^{+}$)-pair in its singlet state)
are related to each other by the following approximate equation
\begin{equation}
 \Gamma_{4 \gamma}({\rm Ps}) \approx 0.274 \Bigl(\frac{\alpha}{\pi}\Bigr)^2
 \Gamma_{2 \gamma}({\rm Ps}) \label{e4}
\end{equation}
Since the three-body Ps$^-$ ion contains only one singlet electron-positron
pair, then from Eq.(\ref{e4}) one finds an analogous expression for the
Ps$^-$ ion
\begin{equation}
 \Gamma_{4 \gamma}({\rm Ps}^-) \approx 0.274
 \Bigl(\frac{\alpha}{\pi}\Bigr)^2 \Gamma_{2 \gamma}({\rm Ps}^-) \label{e4a}
\end{equation}
where $\Gamma_{4 \gamma}$(Ps$^-)$ and $\Gamma_{2 \gamma}$(Ps$^-)$ are the
corresponding annihilation rates of the Ps$^-$ ion. For the two-photon
annihilation rate $\Gamma_{2 \gamma}$ in Eq.(\ref{e4a}) one can use the
explicit expression Eq.(\ref{An2g}). Formally, in Eq.(\ref{e4a}) the formula
for the $\Gamma_{2 \gamma}$ rate must be used which does not contain the
lowest order radiative correction. But, for approximate evaluations we can
ignore such a small difference in $\Gamma_{2 \gamma}$. For the five-photon
annihilation rate in the Ps$^-$ ion one analogously finds the following
result
\begin{equation}
 \Gamma_{5 \gamma}({\rm Ps}^-) \approx 0.177
 \Bigl(\frac{\alpha}{\pi}\Bigr)^2 \Gamma_{3 \gamma}({\rm Ps}^-) \label{e5a}
\end{equation}
This result is based on the formula from \cite{PRA83}.

By using the formulae Eq.(\ref{e4a}) and Eq.(\ref{e5a}) and the numerical
values of the $\Gamma_{2 \gamma}$ and $\Gamma_{3 \gamma}$ annihilation rates
computed above we can evaluate the four- and five-photon annihilation rates
in the Ps$^-$ ion. The corresponding numerical values are $\Gamma_{1
\gamma} \approx$ 3.82491$\cdot 10^{-2}$ $sec^{-1}, \Gamma_{2 \gamma}
\approx$ 2.08048530525$\cdot 10^{9}$ $sec^{-1}, \Gamma_{3 \gamma} \approx$
5.6364151550$\cdot 10^{6}$ $sec^{-1}, \Gamma_{4 \gamma} \approx$ 3.075$\cdot
10^{3}$ $sec^{-1}$ and $\Gamma_{5 \gamma} \approx 5.383$ $sec^{-1}$. The
$\Gamma_{4 \gamma}$ and $\Gamma_{5 \gamma}$ rates have not been evaluated
in earlier studies. The computed annihilation rates $\Gamma_{2 \gamma},
\Gamma_{3 \gamma}, \Gamma_{4 \gamma}, \Gamma_{5 \gamma}$ and $\Gamma_{1
\gamma}$ allow one to determine the effective life-time of the Ps$^-$ ion.
The numerical computation of these annihilation rates performed in this
Section essentially solves the problem of positron annihilation in the
Ps$^-$ ion. It is interesting to note that the $\Gamma_{4 \gamma}$ and
$\Gamma_{5 \gamma}$ annihilation rates are very small in value. By using the
numerical values of all computed annihilation rates one finds that in the
Ps$^-$ ion, $\Gamma_{3 \gamma} \gg \Gamma_{4 \gamma} \gg \Gamma_{5 \gamma}
\gg \Gamma_{1 \gamma}$. Formally, this relationship means that the Ps$^-$
ion can essentially be described as a system with the two annihilation rates
($\Gamma_{2 \gamma}$ and $\Gamma_{3 \gamma}$) only. The partial annihilation
probabilities are:
\begin{equation}
 \Delta_{2 \gamma} = \frac{\Gamma_{2 \gamma}}{\Gamma_{2 \gamma} + \Gamma_{3
 \gamma}} \approx 0.9973 \; \; \; , \; \; \;
 \Delta_{3 \gamma} = \frac{\Gamma_{3 \gamma}}{\Gamma_{2 \gamma} + \Gamma_{3
 \gamma}} \approx 0.0027 .
\end{equation}
In other words, $\approx$ 99.7 \% of all Ps$^-$ ions decay with the emission
of two photons, while $\approx$ 0.3 \% of these ions decay with the emission
of three photons. It is interesting to note that the same situation can be
found in other polyelectrons, including bi-positronium Ps$_2$,
bi-positronium ion Ps$_2 e^{-}$, etc. Each of these systems has only two
annihilation channels: (a) two-photon channel and (b) three-photon channel.
The rest of the annihilation channels can be ignored for all polyelectrons.
This allows one to describe the annihilation of arbitrary polyelectrons and
their mixtures.

\section{Annihilation of the bi-positronium}

Annihilation of the $(e^{-}, e^{+})$-pair in the four-body bi-positronium
Ps$_2$ can proceed with the emission of one-, two-, three- and larger number
of photons; yet in the bi-positronium Ps$_2$ the zero-photon
annihilation is also possible. Let us present the known analytical formulae
for the partial annihilation rates $\Gamma_{n \gamma}$ in bi-positronium
Ps$_2$. First, consider the two-photon annihilation. The analytical
expression for the two-photon annihilation rate which also includes the
lowest-order radiative correction \cite{Harr} takes the form
\begin{eqnarray}
 \Gamma_{2 \gamma} = \pi \Bigl[ 1 - \frac{\alpha}{\pi} \Bigl(5 -
 \frac{\pi^2}{4} \Bigr) \Bigr] \cdot \alpha^4 \cdot n (c a_0^{-1}) \cdot
 \langle \delta_{+-} \rangle = 50.17280268904 \cdot 10^9 \cdot n \cdot
 \langle \delta_{+-} \rangle \; \; \; sec^{-1} , \label{e2g4}
\end{eqnarray}
where $n$ is the total number of electron-positron pairs in polyelectron.
For bi-positronium Ps$_2$ one finds $n = 2 \cdot 2 = 4$. In Eq.(\ref{e2g4})
the notation $\langle \delta_{+-} \rangle$ designates the expectation value
of the electron-positron delta-function $\delta_{+-}$. It is assumed
everywhere in this study that all few-particle wave functions are properly
antisymmetrized upon identical particles. Now, by using our best expectation
value obtained for the electron-positron delta-function $\langle \delta_{+-}
\rangle \approx 2.211775 \cdot 10^{-2}$ (our method was described in
\cite{FrBa}) one finds that $\Gamma_{2 \gamma}($Ps$_2) \approx 4.4385952
\cdot 10^{9}$ $sec^{-1}$. The non-relativistic energy of the ground state of
the bi-positronium Ps$_2$ obtained with the same wave function is -0.5160
0379 0316 $a.u.$

The three-photon annihilation rate $\Gamma_{3 \gamma}($Ps$_2)$ can be
written in the form
\begin{eqnarray}
 \Gamma_{3 \gamma} = \frac{4}{3} (\pi^2 - 9) \alpha^5 \cdot n (c a_0^{-1})
 \cdot \langle \delta_{+-} \rangle = 1.35927298 \cdot 10^7 \cdot n \cdot
 \langle \delta_{+-} \rangle \; \; \; sec^{-1} . \label{e3g4}
\end{eqnarray}
By using the same expectation value of $\langle \delta_{+-} \rangle$ one
finds that $\Gamma_{3 \gamma}($Ps$_2) \approx 1.202497 \cdot 10^7$
$sec^{-1}$.

The one-photon $(e^{-}, e^{+})$-pair annihilation in the bi-positronium
Ps$_2$ proceeds with the emission of one fast electron or positron.
Formally, in the bi-positronium we have two one-photon annihilation rates
$\Gamma_{1 \gamma}(e^{-})$ and $\Gamma_{1 \gamma}(e^{+})$, where the
notation $e^{-}/e^{+}$ means that the fast electron/positron is emitted
during the one-photon annihilation. In actual applications we can always
assume that $\Gamma_{1 \gamma}(e^{-}) = \Gamma_{1 \gamma}(e^{+})$. In this
case the total one-photon annihilation rate $\Gamma_1  = \Gamma_{1
\gamma}(e^{-}) + \Gamma_{1 \gamma}(e^{+}) = 2 \Gamma_{1 \gamma}(e^{-})$ for
the Ps$_2$ system is
\begin{eqnarray}
 \Gamma_{1 \gamma} = \frac{128 \pi^2}{27} \cdot \alpha^{8}
 (c a_0^{-1}) \cdot \langle \delta_{+--} \rangle = 2.1315138 \cdot 10^4
 \cdot \langle \delta_{+--} \rangle \; \; \; sec^{-1} ,
\end{eqnarray}
where $\langle \delta_{+--} \rangle = \langle \delta_{++-} \rangle$ in the
Ps$_2$ system. By using our best-to-date numerical value for the $\langle
\delta_{+--} \rangle$ expectation value ($\langle \delta_{+--} \rangle
\approx 9.11034 \cdot 10^{-5}$) we can evaluate the one-photon annihilation
rate in the bi-positronium Ps$_2$ as $\approx 1.94188 \cdot 10^{-1}$
$sec^{-1}$. The analytical expression for the zero-photon annihilation rate
in Ps$_2$ is
\begin{eqnarray}
 \Gamma_{0 \gamma} = \frac{147 \sqrt{3} \pi^3}{2} \cdot \alpha^{12}
 (c a_0^{-1}) \cdot \langle \delta_{++--} \rangle =
 5.0991890 \cdot 10^{-4} \cdot \langle \delta_{++--} \rangle \; \; \;
 sec^{-1}
\end{eqnarray}
where $\langle \delta_{++--} \rangle$ is the expectation value of the
four-particle delta-function in the bi-positronium Ps$_2$. Now, by using the
$\langle \delta_{++--} \rangle \approx 4.5614 \cdot 10^{-6}$ expectation
value one finds that $\Gamma_{0 \gamma}$(Ps$_2) \approx 2.32197 \cdot
10^{-9}$ $sec^{-1}$.

Consider the cases of four- and five-photon annihilation in the
bi-positronium Ps$_2$. Based on the results from \cite{PRA83} we can write
the corresponding annihilation rates in the form
\begin{equation}
 \Gamma_{4 \gamma}({\rm Ps}_2) \approx 0.274
 \Bigl(\frac{\alpha}{\pi}\Bigr)^2 \Gamma_{2 \gamma}({\rm Ps}_2) \label{e4a4}
\end{equation}
and
\begin{equation}
 \Gamma_{5 \gamma}({\rm Ps}_2) \approx 0.177
 \Bigl(\frac{\alpha}{\pi}\Bigr)^2 \Gamma_{3 \gamma}({\rm Ps}_2) ,
 \label{e54}
\end{equation}
where $\Gamma_{4 \gamma}($Ps$_2)$ and $\Gamma_{5 \gamma}($Ps$_2)$ are the
four- and five-photon annihilation rates of bi-positronium Ps$_2$. The
expressions for the two- and three-photon annihilation rates are given by
Eq.(\ref{e2g4}) and Eq.(\ref{e3g4}), respectively. The numerical values
of these annihilation rates for the Ps$_2$ system are $\Gamma_{0 \gamma}
\approx$ 2.32197$\cdot 10^{-9}$ $sec^{-1}, \Gamma_{1 \gamma} \approx$
1.94188$\cdot 10^{-1}$ $sec^{-1}, \Gamma_{2 \gamma} \approx$ 4.4385952$\cdot
10^{9}$ $sec^{-1}, \Gamma_{3 \gamma} \approx$ 1.202497$ \cdot 10^7$
$sec^{-1}, \Gamma_{4 \gamma} \approx$ 6.562$\cdot 10^{3}$ $sec^{-1}$ and
$\Gamma_{5 \gamma} \approx$ 11.484 $sec^{-1}$. The $\Gamma_{4 \gamma}$ and
$\Gamma_{5 \gamma}$ rates have not been calculated in earlier studies.
The numerical computation of these annihilation rates practically solves the
annihilation problem for the bi-positronium system Ps$_2$. A few
annihilation rates, e.g., $\Gamma_{6 \gamma}$ and $\Gamma_{7 \gamma}$, which
still remain unknown, are very small. Further, they are not of any interest
in current applications. Note that in the first (and very accurate)
approximation the annihilation of the Ps$_2$ system can be described with
the use of the two annihilation rates ($\Gamma_{2 \gamma}$ and $\Gamma_{3
\gamma}$) only.

\section{Conclusion}

We have considered annihilation of the electron-positron pairs in the
three-body ion Ps$^-$ and four-body bi-positronium `molecule' Ps$_2$. The
rates of the two-, three-, four- and five-photon annihilation have been
determined numerically for the ground states in each of these systems. In
our calculations we have used the expectation values of delta-functions
determined with the help of highly accurate wave functions. The
corresponding numerical values of these annihilation rates (and also
$\Gamma_{1 \gamma}$ and $\Gamma_{0 \gamma}$ annihilation rates) for the
Ps$^-$ ion and Ps$_2$ `molecule' can be found in the text and Abstract. With
the final computation of these annihilation rates we have to note that this
work concludes many years of intense research of the problem of annihilation
of electron-positron pairs from the bound (ground) states of the Ps$^-$ ion
and Ps$_2$ system. As follows from the results of this work, in the first
(and very good) approximation annihilation in the Ps$^-$ ion and Ps$_2$
system can be considered by taking into account the two largest annihilation
rates only, i.e. $\Gamma_{2 \gamma}$ and $\Gamma_{3 \gamma}$. This
conclusion can be generalized to arbitrary polyelectrons (polyleptons).

Note that the three-body Ps$^-$ ion has been created in the laboratory by
Mills \cite{Mil1}, \cite{Mil2}. The total annihilation rate of the Ps$^-$
ion has been measured in \cite{Mil2}, where it was found that $\Gamma$
(Ps$^-) \approx 2.09 \cdot 10^{9}$ $sec^{-1}$. Annihilation rates in the
bi-positronium Ps$_2$ have never been measured experimentally. In fact, an
isolated bi-positronium `molecule' Ps$_2$ has not been created in the
laboratory. Nevertheless, the bi-positronium is of great interest in
astrophysics, solid state physics and other problems (see discussion and
references in \cite{FrBa}).

\begin{center}
 {\bf Acknowledgement}
\end{center}

I would like to thank Peter G. Komorowski for useful discussions and help
with the manuscript.

\end{document}